\documentclass[12pt, a4paper]{article} 
\usepackage[utf8]{inputenc}                  

\usepackage{geometry}    
\usepackage{authblk} 
\usepackage[round]{natbib} 
\usepackage{hyperref}
\hypersetup{citecolor=green, urlcolor=cyan}

\usepackage{lineno} 

\usepackage{graphicx} 
\usepackage{epstopdf}
\usepackage{color} 
\usepackage[title]{appendix}
\usepackage[parfill]{parskip} 

\usepackage{mathptmx}
\usepackage{booktabs} 
\usepackage{rotating}
\usepackage{amsmath}
\usepackage{amsfonts}
\usepackage{array} 
\usepackage{paralist} 
\usepackage{verbatim} 
\usepackage{subfig} 
\usepackage[x11names,svgnames,dvipsnames]{xcolor}

\usepackage{fancyhdr} 
\usepackage{sectsty}
\usepackage{sectsty}
\allsectionsfont{\rmfamily\bfseries} 
\subsectionfont{\rmfamily\bfseries\itshape}
\subsubsectionfont{\rmfamily\itshape}

\newcolumntype{L}[1]{>{\raggedright\let\newline\\\arraybackslash\hspace{0pt}}m{#1}}

\usepackage[nottoc,notlof,notlot]{tocbibind} 
\usepackage[titles,subfigure]{tocloft} 

\newcommand*\patchAmsMathEnvironmentForLineno[1]{%
  \expandafter\let\csname old#1\expandafter\endcsname\csname #1\endcsname
  \expandafter\let\csname oldend#1\expandafter\endcsname\csname end#1\endcsname
  \renewenvironment{#1}%
     {\linenomath\csname old#1\endcsname}%
     {\csname oldend#1\endcsname\endlinenomath}}%
\newcommand*\patchBothAmsMathEnvironmentsForLineno[1]{%
  \patchAmsMathEnvironmentForLineno{#1}%
  \patchAmsMathEnvironmentForLineno{#1*}}%
\AtBeginDocument{%
\patchBothAmsMathEnvironmentsForLineno{equation}%
\patchBothAmsMathEnvironmentsForLineno{align}%
\patchBothAmsMathEnvironmentsForLineno{flalign}%
\patchBothAmsMathEnvironmentsForLineno{alignat}%
\patchBothAmsMathEnvironmentsForLineno{gather}%
\patchBothAmsMathEnvironmentsForLineno{multline}%
}

\providecommand{\keywords}[1]
{
  \small	
  \textbf{\textit{Keywords---}} #1
}

\begin{document}

\title{Assimilation of semi-qualitative sea ice thickness data with the EnKF-SQ}
\author[1,\thanks{Corresponding author (abhishek.shah@nersc.no)}]{Abhishek~Shah}
\author[1]{Laurent~Bertino}
\author[1]{Fran\c cois~Counillon}
\author[2]{Mohamad~El Gharamti}
\author[1]{Jiping~Xie}

\affil[1]{Nansen Environmental and Remote Sensing Center, Bergen, Norway}
\affil[2]{National Center for Atmospheric Research, Boulder, Colorado}

\date{}

\maketitle
\begin{abstract}
A newly introduced stochastic data assimilation method, the Ensemble Kalman Filter Semi-Qualitative (EnKF-SQ) is applied to a realistic coupled ice-ocean model of the Arctic, the TOPAZ4 configuration, in a twin experiment framework. The method is shown to add value to range-limited thin ice thickness measurements, as obtained from passive microwave remote sensing, with respect to more trivial solutions like neglecting the out-of-range values or assimilating climatology instead. 

Some known properties inherent to the EnKF-SQ are evaluated: the tendency to draw the solution closer to the thickness threshold, the skewness of the resulting analysis ensemble and the potential appearance of outliers. The experiments show that none of these properties prove deleterious in light of the other sub-optimal characters of the sea ice data assimilation system used here (non-linearities, non-Gaussian variables, lack of strong coupling). The EnKF-SQ has a single tuning parameter that is adjusted for best performance of the system at hand. The sensitivity tests reveal that the results do not depend critically on the choice of this tuning parameter. The EnKF-SQ makes overall a valid approach for assimilating semi-qualitative observations into high-dimensional nonlinear systems. 
\end{abstract}


\section{Introduction}
Sea ice plays a crucial role in the Arctic climate as it modulates the exchange of heat and moisture between the ocean and the atmosphere \citep{Aagaard1989,Screen2010}. Different studies have shown that accurate knowledge of the Sea Ice Thickness (SIT) is beneficial for the Arctic sea ice predictability \citep{Day2014,Collow2015,Guemas2014}. The SIT observations from the European Space Agency (ESA) Soil Moisture and Ocean Salinity (SMOS) mission are available in near-real time, at daily frequency during the cold season (October-April). The retrieval method for SMOS SIT observations is based on measurements of the brightness temperature at a low frequency microwave (1.4 GHz, L-band: wavelength of $21$ cm) \citep{Kaleschke2010}. The representative depth for the L-Band microwave frequency into the sea ice is about $0.5$ m for first-year level ice \citep{Kaleschke2010,Huntemann2014}. Few studies have shown that assimilating thin SIT from SMOS into coupled ice-ocean model, using ensemble based Data Assimilation (DA) techniques, is able to improve the SIT forecast without being detrimental to other properties \citep[e.g.,][]{Yang2014, Xie2016, Fritzner2019}. All of these studies, however, ignore the saturated observations of thick ice. 

Measurements of thick sea ice on basin-wide scales are also available from laser altimeters onboard ICESat \citep{Forsberg2005} or from radar altimeters on the European Remote Sensing (ERS), Envisat, CryoSat-2 and Sentinel-3 \citep{Connor2009,Laxon2013,Ricker2014}. CryoSat-2 SIT is provided in near-real time \citep{Tilling2016} but still contains considerable large uncertainties caused by the lack of auxiliary data on snow depth. These uncertainties are proportionally larger for thin ice (i.e., $<1$ m) and hence CryoSat-2 practically measures thick sea ice only. A merged product of weekly SIT observations in the Arctic from the CryoSat-2 altimeter and SMOS radiometer, referred as CS2SMOS, has also been developed by combining the two complementary datasets \citep{Kaleschke2015,Ricker2017} and made available during the winter months since October 2010. However, the combination of the two satellites is not perfect as biases have been revealed on overlapping areas \citep{Wang2016, Ricker2017}. Recently, \cite{Xie2018} successfully assimilated the merged SIT product CS2SMOS into the TOPAZ4 coupled ocean-sea ice reanalysis system \citep{Sakov2012} for the Arctic.

While assimilating a merged SIT map, rather than two satellites data streams is practically convenient, the uncertainty of the merged data is more difficult to quantify and bad quantification of the uncertainty may affect the assimilation performance negatively \citep{Mu2018}\footnote{It should be noted that the comparison of assimilating merged versus separate data is not informative because their observation errors are not equivalent}. The ability to use well-justified observation errors in data assimilation is sufficiently important to motivate the assimilation of the two separate SIT data streams rather than one merged product. This implies that their detection limits should be taken into account by the data assimilation method. 

In DA, observations are used to reduce the error of the state variables so that the forecast skill can be enhanced. Many observations can only be retrieved within a limited interval of the values that the observed quantity would take in nature. In other words, observations may have a detection limit. One such example is the aforementioned observation of SIT from SMOS. Although, the SIT observations with detection limit do not provide quantifiable data (hard data) above its detection limit, they do give qualitative information (soft data). For instance, the ice could be thicker than a known threshold. Studies from \cite{Shah2018} and \cite{Borup2015} have shown that assimilating soft data with linear and non-linear toy models using ensemble-based DA methods have the potential to improve the accuracy of the forecast. Therefore, not considering soft data in the assimilation procedure is a potential loss of meaningful information. 

Assimilating only thin ice observations, as in \citet[Figure 5 and 6]{Xie2016}, induces a low bias, which is caused by the partial nature of the observation of thin ice. With a new method intended for semi-qualitative data as the EnKF-SQ, the question arise whether this bias can be mitigated or not? The comparison of the EnKF-SQ to the perfect Bayesian solution \citep{Shah2018} shows that the EnKF-SQ analysis does not coincide with the Bayesian posterior and bears inherent biases: in the case of hard data, the Bayesian and EnKF-SQ posteriors are nearly the same. However, for out-of-range observations and mode of a prior within the observable range, only the maximum likelihood of the EnKF-SQ analysis is preserved but its distribution is flatter than the Bayesian solution with a thicker tail in the unobservable range, so the expectation is too high. Based on this, the EnKF-SQ is expected to be unbiased for thin SIT observations. Nevertheless, it should show a positive bias for out-of-range observations. Further, the thicker tail of the EnKF-SQ analysis distribution in the unobservable range makes it relatively skewed, which is undesirable in a Kalman filtering context.

In this study, we implement and test the overall performance of the stochastic ensemble Kalman filter semi-qualitative (EnKF-SQ) \citep{Shah2018} in a twin experiment where synthetic SMOS-like SIT observations, with an upper detection limit, are assimilated into a coupled ocean-sea ice forecasting system. The objective is to test the potential of the EnKF-SQ for assimilating soft data with a state of the art ocean and sea ice prediction system, namely TOPAZ4. In addition, a number of single-cycle assimilation experiments using the EnKF-SQ are performed to investigate the sensitivity to the ensemble size and out-of-range observation uncertainty.  

This paper is organized as follows: Section \ref{sec:TOPAZ_system} introduces the main components of the TOPAZ4 system including the model and the EnKF-SQ DA scheme used in the assimilation experiments. In Section \ref{sec:exper-setup}, the synthetic ice thickness data are outlined together with the assimilation setup. Section \ref{sec:assim_results} discusses the results of the various assimilation experiments. A general discussion of the study concludes the paper in Section \ref{sec:conclusion}.

\section{The TOPAZ system} \label{sec:TOPAZ_system}
\subsection{Model setup}\label{subsec:Model system}
The ocean general circulation model used in the TOPAZ4 system is the version 2.2 of the Hybrid Coordinate Ocean Model (HYCOM) developed at the University of Miami \citep{Bleck2002,Chassignet2003}. The TOPAZ4 implementation of HYCOM uses hybrid coordinates in the vertical, which smoothly shift from isopycnal layers in the stratified open ocean to \textit{z} level coordinates in the unstratified surface mixed layer. 

The HYCOM ocean model is coupled to a one-thickness category sea ice model. The single ice thickness category thermodynamics are described in \cite{Drange1996} and the ice dynamics use the Elastic-Viscous-Plastic (EVP) rheology of \cite{Hunke1997} with a modification from \cite{Bouillon2013}. The momentum exchange between the ice and the ocean is given by quadratic drag formulas. The model has a minimum thickness of $10$ cm for both new and melting ice. 

\begin{figure}[h]
\includegraphics[scale=0.3]{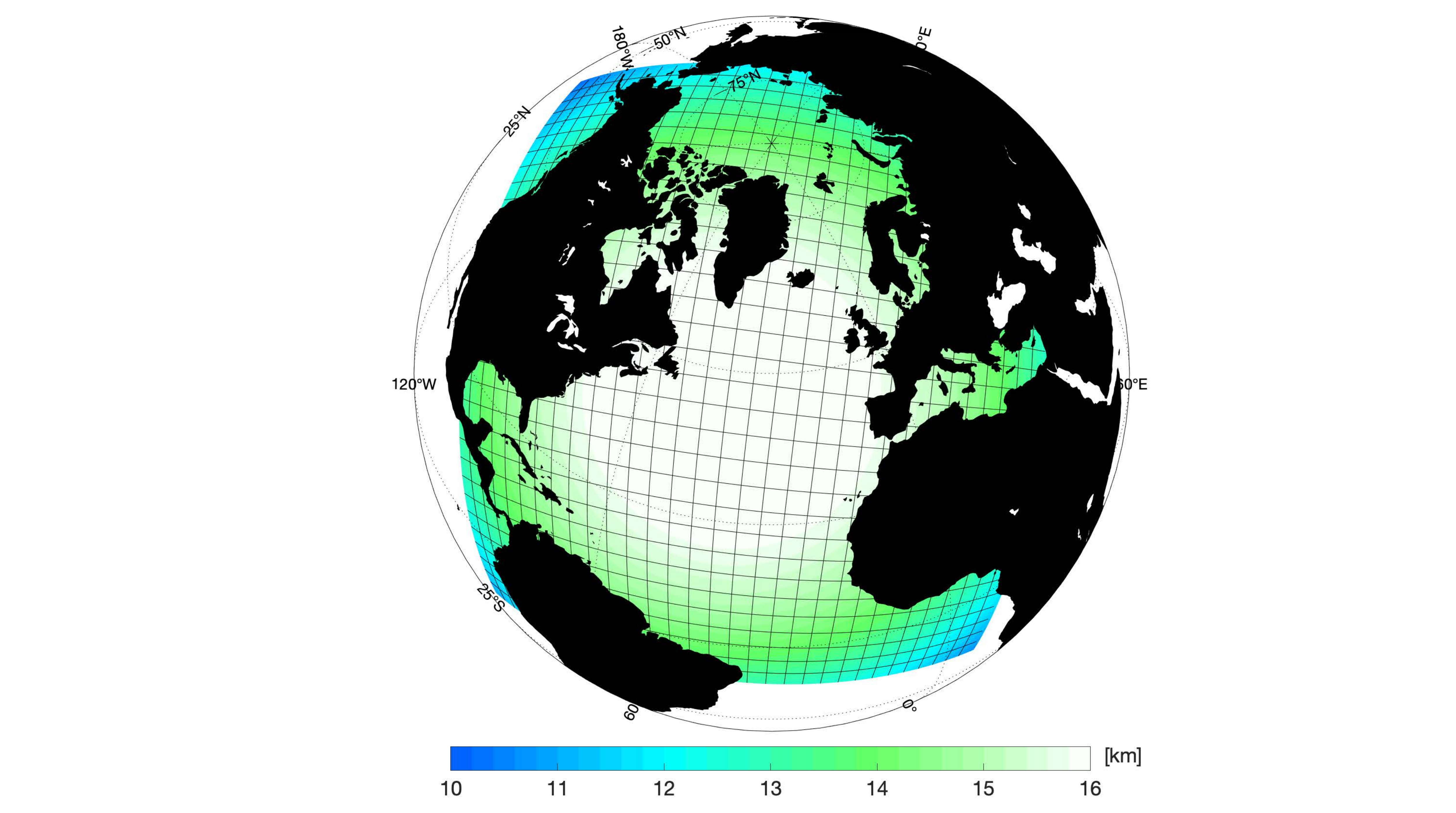}
\begin{center}
\caption{The TOPAZ4 model domain. Background color shading shows the horizontal grid resolution (km) while solid black color represents land. \label{fig:region_of_focus} }
\end{center}
\end{figure}

The model domain covers the North Atlantic and Arctic basins as shown in Figure~\ref{fig:region_of_focus}. The model grid is created with conformal mapping \citep{Bentsen1999} and has a quasi-homogeneous horizontal resolution between $12-16$ km in the whole domain. The grid has $880\times800$ horizontal grid points.

\subsection{The Ensemble Kalman Filter Semi-Qualitative, EnKF-SQ} \label{subsec:The EnKF-SQ}

The EnKF-SQ \citep{Shah2018} uses an ensemble of model states to estimate the error statistics closely following the stochastic EnKF algorithm \citep{Burgers1998, Evensen2004}. The stochastic EnKF is a two-step filtering method alternating forecast and analysis steps. In the forecast step, the ensemble of model states is integrated forward in time and when observations become available, an analysis of every forecast member, $\mathbf{x}_{i}^{f}$ for $i\in{1,2,...,N}$, is computed as follows:
\begin{equation}
    \mathbf{x}_{i}^{a}=\mathbf{x}_{i}^{f}+\mathbf{K}(\mathbf{y}_{i}-\mathbf{H}\mathbf{x}_{i}^{f}),\label{eq:EnKFSQ_hard_update_eq}
\end{equation}
\begin{equation} 
\mathbf{K}=\mathbf{P}^{f}\mathbf{H}^{T}(\mathbf{H}\mathbf{P}^{f}\mathbf{H}^{T}+\mathbf{R})^{-1},\label{eq:Kalman_inrange} 
\end{equation}
where $\mathbf{K}$ is the Kalman gain matrix; $\mathbf{x}_{i}$ is the $i^{th}$ ensemble state member; $\mathbf{H}$ is the observation operator, mapping the state variable to the observation space (could be non-linear); $\mathbf{R}$ is the observation error covariance matrix; $\mathbf{y}_{i}$ is the $i^{th}$ perturbed observation vector generated from  ${\cal N}(\mathbf{y},\mathbf{R})$ and $\mathbf{P}^{f}$ is the ensemble forecast error covariance matrix. The superscripts $a$, $f$, and $T$ stand for analysis, forecast, and matrix transpose, respectively. In practice, $\mathbf{P}^{f}$ is never computed explicitly and is instead decomposed as follows:
\begin{equation} 
\mathbf{P}^{f}=\frac{1}{N-1}\sum\limits_{i=1}^N(\mathbf{x}_{i}^{f}-\bar{\mathbf{x}}^{f})(\mathbf{x}_{i}^{f}-\bar{\mathbf{x}}^{f})^{T},
\end{equation}
where $\bar{\mathbf{x}}^{f}$ is the mean of the forecast ensemble.

The EnKF-SQ is intended to explicitly assimilate observations with a detection limit. These are divided into two categories depending on whether they are within or outside the observable range. If the observed quantity is within it, the quantitative (hard) data is assimilated as in the stochastic EnKF, otherwise it is considered a qualitative (soft) data and treated differently. 

The specific value and error statistics of the out-of-range (OR) observations are unknown. In order to assimilate OR observations, an assumption needs to be made about its likelihood. Following \cite{Shah2018}, a virtual observation is created at the detection limit and then a two-piece Gaussian observation likelihood is constructed around it. A two-piece Gaussian distribution is obtained by merging two opposite halves of two different Gaussian probability density functions (pdfs) at their common mode, given as follows: 
\begin{equation}
f(x)=\begin{cases}
we^{-{\left(x-\mu\right)^{2}}/{2\sigma_{ir}^{2}}},\hspace{0.5cm}x{\leq}\mu,\\
we^{-{\left(x-\mu\right)^{2}}/{2\sigma_{or}^{2}}},\hspace{0.5cm}x{>}\mu,
\end{cases} \label{eq:2_piece}
\end{equation}
where $w=\sqrt{\frac{2}{\pi}}(\sigma_{ir}+\sigma_{or})^{-1}$ is a normalizing constant, $\mu$ is the detection limit and also the common mode of two different normal distribution; $\sigma_{ir}$ and $\sigma_{or}$ are in-range and OR observation error standard deviations (std), respectively.

\begin{figure}[h]
\begin{center}
\includegraphics[scale=0.3]{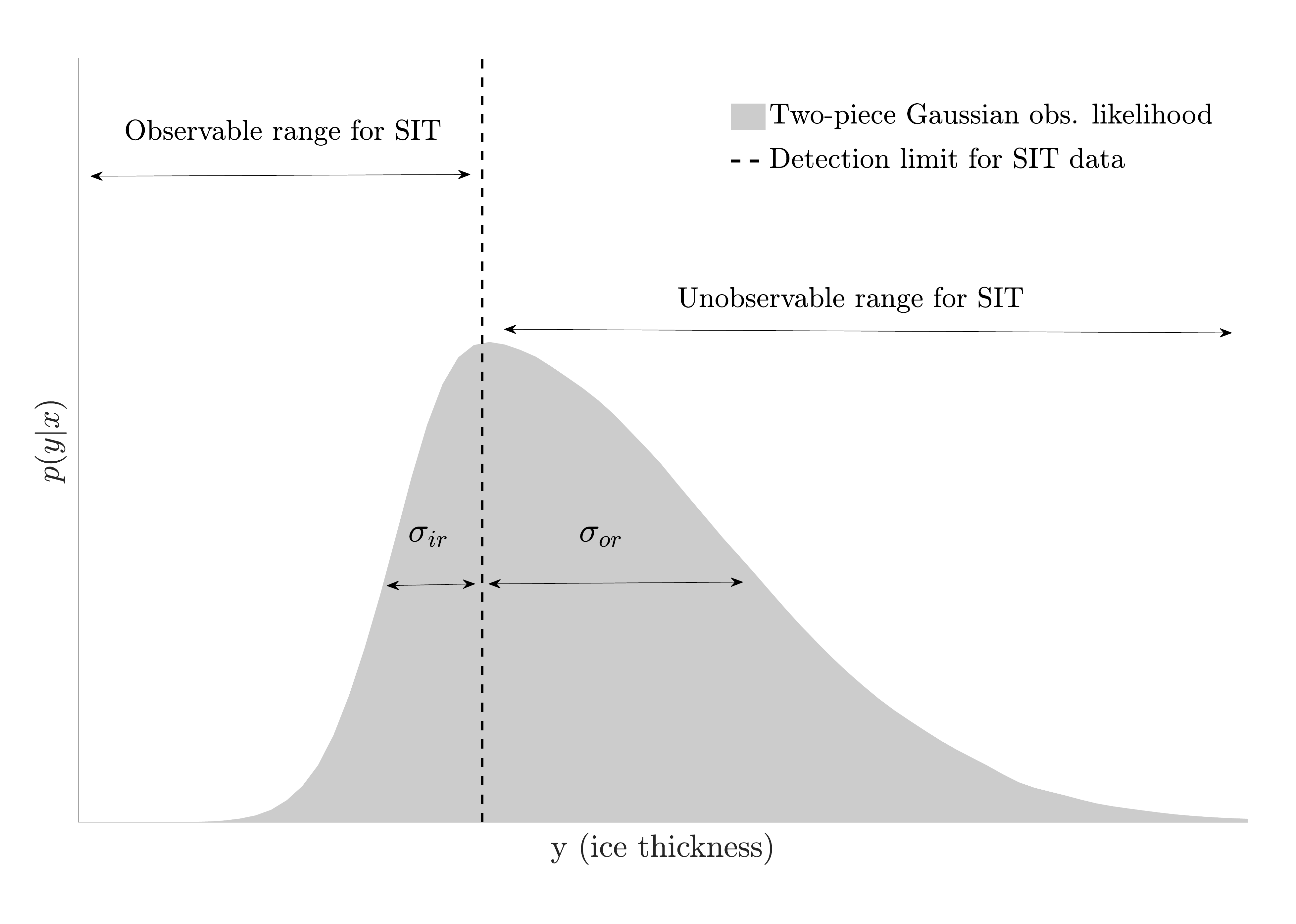}
\caption{Illustration of the two-piece Gaussian OR-observation likelihood for SMOS-like thin SIT. $\sigma_{ir}$ is an in-range and $\sigma_{or}$ is the out-of-range observation error standard deviations, respectively. \label{fig:Illustration_2_piece_Gaussian_likelihood} }
\end{center}
\end{figure}

Figure \ref{fig:Illustration_2_piece_Gaussian_likelihood} is an illustration of a two-piece Gaussian observation likelihood for OR SIT observations. On the left hand side of the detection limit, it is assumed that $\sigma_{ir}$, inside the observable range, is defined by the observation uncertainty of hard data at the detection limit. An observation could possibly fall outside the detection limit, due to observation errors, even though its true value is within the observable range. On the right hand side, it is assumed that the $\sigma_{or}$ (eq. \ref{eq:sig_or}) in the unobservable range is defined with the help of a climatological mean for SIT above the detection limit so that extremely high values, which are usually less realistic, receive a lower likelihood \citep{Shah2018}.  
\begin{equation}
\sigma_{or}=\underset{\text{Climatological mean}}{\underbrace{\int\displaylimits_{\mu}^{+\infty}yf_c(y)dy}}-\mu. \label{eq:sig_or}
\end{equation}
$f_c(y)$ is the pdf of the climatological data of the observed quantity. The two-piece Gaussian observation likelihood for soft data is denoted, hereafter, by ${\cal{N}}_{2p}(\mu,\sigma_{ir}^{2},\sigma_{or}^{2})$.  The EnKF-SQ pre-processes the observations by sorting them as either hard $y^h$ or soft $y^s$. The observation errors are assumed uncorrelated in space, i.e. $\mathbf{R}$ is diagonal. 

\begin{description}
\item [\textit{Update step of the EnKF-SQ}]
\end{description}
For each forecast member $\mathbf{x}_i^{f}$ ($i \in 1, 2, \hdots, N$):
\begin{enumerate}

    \item For each soft data $y^s_j$, check whether the observed forecast ensemble member is within the observable range or not.
    
    \item If $\mathbf{H}_{j}\mathbf{x}_{i}^{f} \leq \mu$, set observation error variance $\mathbf{R}_{j,j}=\sigma_{ir}^{2}$ otherwise $\mathbf{R}_{j,j}=\sigma_{or}^{2}$ implying that members inside (outside) the observable range are updated with data parameterized using in-range $\sigma_{ir}^{2}$ (out-of-range $\sigma_{or}^{2}$).
    
    \item After looping over all soft data, compute the Kalman gain $\mathbf{K}_{i}$ as in Eq.~\ref{eq:Kalman_inrange} with the updated observation error covariance matrix $\mathbf{R}$. For each $\mathbf{x}_{i}^{f}$, a different Kalman gain $\mathbf{K}_{i}$ is calculated.
    
    \item Evaluate the $i^{\text{th}}$ analysis member $\mathbf{x}_{i}^{a}$ as in Eq.~\ref{eq:EnKFSQ_hard_update_eq} using $\mathbf{K}_{i}$. The perturbed observations are generated by sampling from ${\cal{N}}(y^h_j,\sigma_h^{2})$ and ${\cal{N}}_{2p}(\mu,\sigma_{ir}^{2},\sigma_{or}^{2})$\footnote{$\sigma_{ir}$ is a special case of $\sigma_h$, for hard data at the detection limit.} for $y^h_j$ and $y^s_j$, respectively. $\sigma_h^{2}$ is the observation error variance for $y_j^{h}$.
    
\end{enumerate}
Loop to next member $i$.

Repeating this process for all forecast members yields the analysis ensemble. For a detailed description of the EnKF-SQ the reader is referred to Section 2 of \cite{Shah2018}. 

\section{Experimental Setup}\label{sec:exper-setup}
\subsection{The synthetic Sea Ice Thickness Data} \label{subsec:SIT data}
The synthetic SIT data used in this study is intended to mimic the SIT data from the SMOS mission with an upper detection limit. In order to evaluate the EnKF-SQ method against a perfectly known truth, synthetic observations are generated using the coupled ocean and one-thickness category sea ice model described earlier in Section \ref{subsec:Model system}. A reference \textit{truth} run (also called nature run) is produced by integrating the coupled ocean sea ice model from $1$ January, $2014$ to $31$ December, $2015$ using unperturbed atmospheric forcing from ERA-Interim \citep{Dee2011}. The run is initialized using member number $100$ from the $100$-member ensemble reanalysis of \cite{Xie2017} on $31$ December, $2013$.   

Synthetic SIT data are then generated for the duration of the assimilation experiment from $11$ November $2014$ to $31$ March $2015$ by perturbing the truth with Gaussian noise of zero mean and standard deviation $\sigma_{obs}$; parameterized as:
\begin{equation}
\sigma_{obs} = {0.06t+0.05}, \label{eq.obs_error}
\end{equation}
where $t$ is the truth for ice thickness in meters. The parameterization is chosen such that observation errors increase for thicker ice, which is a general behaviour of positive-valued variables like SIT. The relationship is obtained through regression of the absolute difference of the daily averaged SIT between the reanalysis product \citep{Xie2017} and the aforementioned reference trajectory from the month of December $2014$ to January $2015$. The resulting relationship (not shown here) is linear with a positive slope. SIT observation error represented in Eq.~\ref{eq.obs_error} is also qualitatively in line with those used by \cite{Xie2016} for SMOS data. 

A single upper detection limit of $1$ m is imposed on the generated SIT observations, as an analogous for  saturation of SMOS data in thick sea ice. The SIT observations are assumed available on every grid cell (except along the coastline) and assimilated on a weekly basis. This is a reasonable assumption as SMOS data comes with a resolution of ($\sim12.5$ km), which is also the resolution of the TOPAZ4 system. Model and observation grids are collocated, thus our experiments neglect potential errors due to interpolation, which is out of the scope of this study.

\subsection{Out-of-range SIT Climatology}
A trivial alternative to the EnKF-SQ in the presence of soft data would be to assimilate climatological data as hard data. It is, therefore, worth investigating how beneficial the assimilation of soft data with the EnKF-SQ is compared to assimilating climatology.

An out-of-range, location-dependent, SIT climatology is computed by taking a time average of the truth (described earlier) for SIT above the detection limit in each grid cell. Averaging is done from January $2014$ to December $2015$, a period that includes two summers and two winters and encompasses the assimilation period. Even though the latter takes place in winter, the climatology has a high bias because by construction it only contains SIT above $1$ m. The observation error variance for the climatological value is also location-dependent, equal to the variance of all reference truth values above the detection limit in the same grid cell.

\subsection{Assimilation setup} \label{subsec:assim_setup}
In contrast to earlier TOPAZ4 studies that updated the whole water column variables \citep{Xie2018}, here the state vector $\mathbf{x}$ consists of only two sea ice variables: SIT and sea ice concentration (SIC). This therefore constitutes a case of a weakly coupled assimilation where the ocean is only updated by dynamical re-adjustments from the sea ice updates. \cite{Kimmritz2018}, have shown that while strongly coupled ocean and sea ice is clearly beneficial, weakly coupled DA can still achieve reasonable results.

In the analysis, sampling errors in the forecast error covariance can give rise to spurious correlations between remote grid points, a problem which may become more pronounced for smaller ensemble sizes \citep{Houtekamer1998}. A common practice to counteract sampling errors is to perform local analysis in which variables at each grid cell are updated using only the observations within a radius of influence $r_{o}$ around the grid cell \citep{Houtekamer1998,Evensen2003}. For simplicity, a single closest local observation within $r_o = 300$ km is used here during the analysis. 

In TOPAZ4, model error is introduced by increasing the model spread via perturbing few forcing fields. The perturbations are pseudo-random fields computed in a Fourier space with a decorrelation time-scale of 2 days and horizontal decorrelation length scale of $250$ km, as described in \cite{Evensen2003}. Perturbed variables include air temperature, wind speeds, cloud cover, sea level pressure \citep[Section 3.3]{Sakov2012} and yield curve eccentricity in the EVP rheology \citep[Table 1]{Hunke1997}. In addition, precipitation is also perturbed with log-normal noise and standard deviation of $100$\%. This affects the snowfall when temperatures are below zero. Snow is an important thermal insulator and therefore hampers sea ice growth/melt.

\subsection{Target Benchmarks} \label{subsec:target_benchmark}
The performance of the EnKF-SQ is compared against three different versions of the stochastic EnKF and a Free run, denoted as follows:
\begin{enumerate}
    \item \textbf{EnKF-ALL}: No detection limit is applied on SIT observations thus even thick ice data from the reference run is assimilated. This run acts as an upper bound for performance because it is the only one that assimilates out-of-range observations as hard data with known statistics, which can be seen as \textit{cheating}. 
    \item \textbf{EnKF-CLIM}: The SIT climatology with climatological variance is assimilated instead of hard data.
    \item \textbf{EnKF-IG}: Only hard data is assimilated and soft data is ignored, similar to \citet{Xie2016}. This run is meant to assess the added value of the EnKF-SQ.
    \item \textbf{Free-run}: The Free-run is the average of the $99$ members without DA. It is run with perturbations, contrarily to the aforementioned single-member truth run. 
\end{enumerate}

To evaluate the performance of the different DA methods, we compute the root mean square error (RMSE) of the ensemble mean at time $t$ as:
\begin{equation}
\text{RMSE}_{t}=\sqrt{\frac{1}{n}\sum\limits_{i=1}^n \left( \bar{\mathbf{x}}_{i,t}^f-\mathbf{x}_{i,t}^r \right)^2},
\end{equation}
where $\mathbf{x}^r$ and $\bar{\mathbf{x}}^f$ is the n-dimensional reference (unperturbed truth) and mean of the prior state vector at time $t$, respectively.  We also monitor the average ensemble spread (AES) for each filter, which we calculate at every assimilation cycle as:
\begin{equation}
\text{AES}_{t} = \sqrt{\frac{1}{n}\sum\limits_{i=1}^{n}\sigma_{i,t}^{2}},
\end{equation}
where $\sigma_{i,t}^2$ can either be the prior or posterior ensemble variance at time $t$, respectively.

\subsection{Ensemble size}
\begin{figure}[h]
\begin{center}
\includegraphics[scale=0.28]{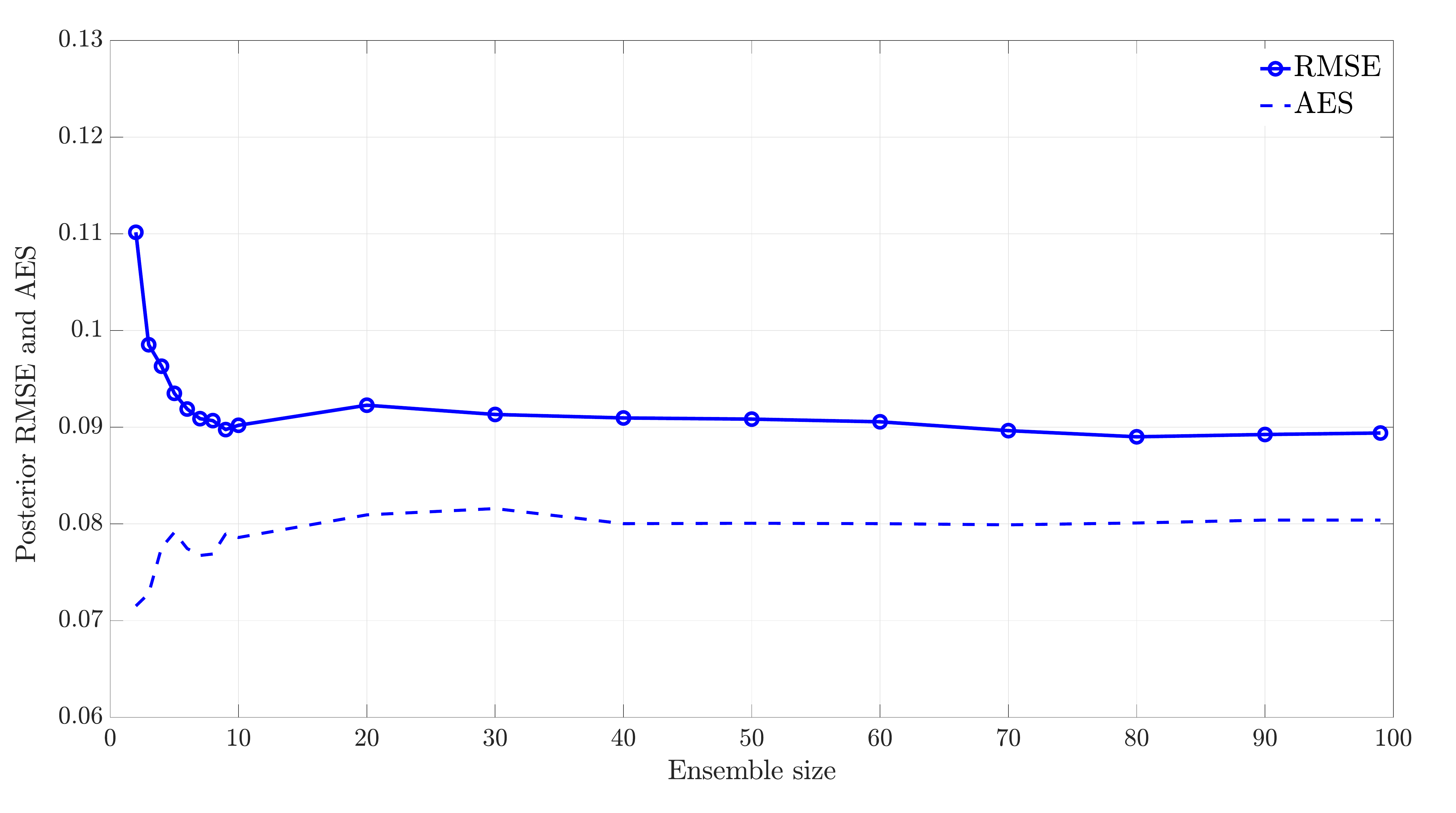}
\caption{Time-averaged posterior RMSE and AES resulting from single cycle assimilation runs for different ensemble sizes using EnKF-SQ. \label{fig:sensitivity_ens_size}}
\end{center}
\end{figure}

In order to select the ensemble size, single-cycle assimilation sensitivity experiments are conducted using EnKF-SQ by varying the ensemble size between $2$ and $99$. The resulting time-averaged RMSE and AES of the posterior SIT estimates are displayed in Figure \ref{fig:sensitivity_ens_size}. The plot indicates that for $N \geq 10$, there is no significant difference in the performance of the EnKF-SQ. This is mostly due to the small size of the local state vector; consisting only of two variables. An ensemble as small as $10$ members is however less likely to succeed on the long term especially if the number of state variable and observations increase. Results from the other three EnKF runs (not shown) showed the exact same behavior. Thus, the initial ensemble is set as the first $99$ members of the reanalysis ensemble of \cite{Xie2016} on $31$ December $2013$. The initial ensemble is then spun up from January, $2014$ until the start of the assimilation experiment (i.e., November $11$) with perturbed forcing to increase the variability. As described earlier, member number $100$ of the reanalysis run was used to generate the truth in this study. 

The assimilation framework is sub-optimal for few reasons, in particular because of the weakly coupled updates. Further, SIT errors are erroneously assumed Gaussian while they are not. These sub-optimalities are not uncommon in realistic applications. They do cause some limited loss of performance but generally do not prevent us from applying the EnKF. 

In terms of computational resources, we used a single processor on supercomputer for each of the four DA methods. The total wall-clock time required by each analysis scheme, to update the SIT and SIC state variables along with the IO operations, is approximately $6$ minutes on a 1.4GHz Cray XE6. This is much less than the TOPAZ4 one-week forward model run, for which each member runs on $134$ parallel processors in approximately $5$ minutes.

\section{Assimilation Results} \label{sec:assim_results}
\subsection{Tuning the EnKF-SQ out-of-range likelihood} \label{subsec:tuning_sq}
The out-of-range standard deviation $\sigma_{or}$ is the only new parameter introduced into the EnKF-SQ compared to the stochastic EnKF. Therefore, it is important to study how the uncertainty in the estimate of $\sigma_{or}$ affects the performance of the EnKF-SQ scheme. For this, we carried out a number of single-cycle assimilation experiments by introducing a scalar multiplier $\alpha$ to equation \ref{eq:sig_or} such that $\sigma_{or}^{*}=\alpha \cdot \sigma_{or}$.

\begin{figure}[h]
\begin{center}
\includegraphics[scale=0.28]{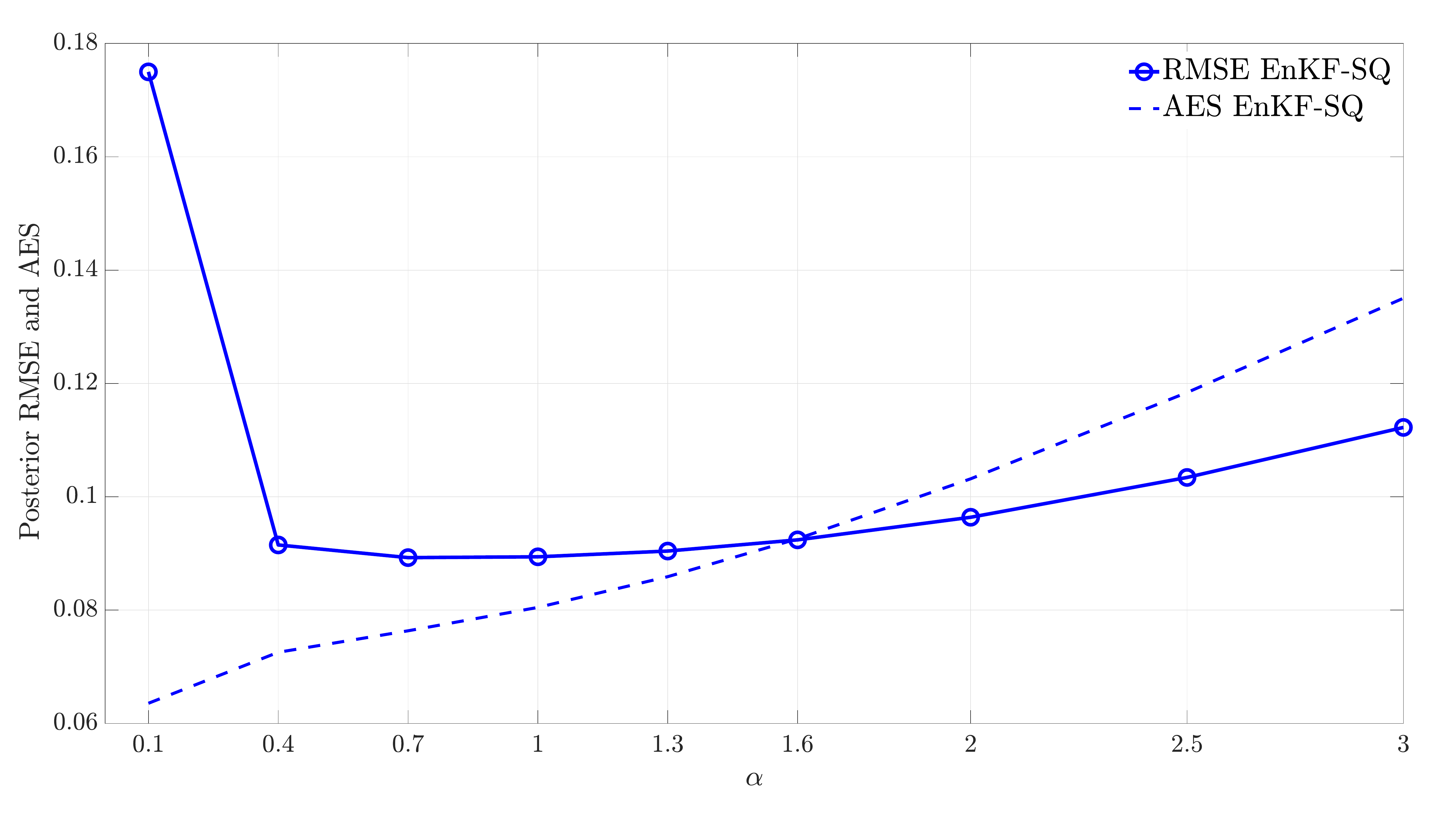}
\caption{Time-averaged posterior RMSE and AES resulting from single cycle assimilation runs for a wide range of the multiplicative factor $\alpha$. \label{fig:sensitivity_alpha}}
\end{center}
\end{figure}

RMSE and AES of the posterior SIT estimates are plotted in Figure \ref{fig:sensitivity_alpha} for a wide range of $\alpha$, varying between $0.1$ and $3.0$. Such a range is very broad for most realistic applications. $\alpha < 0.4$ strongly degrades the accuracy of the EnKF-SQ along with significant decrease in the AES. The large difference between RMSE and AES values, indicate a possible filter divergence. This is because for small $\alpha$ values, the sampling of a two-piece Gaussian likelihood for observation perturbations is prone to generate samples concentrated around the detection limit, thus pulling the analysis close to the detection limit, subsequently reducing the ensemble spread and increasing the RMSE. As $\alpha$ approaches $1$, the RMSE attains the minimum value and further becomes consistent with the AES. When $\alpha$ increases beyond $2$, the sampling of OR likelihood starts producing large perturbations, which makes the analysis increment capricious and eventually deteriorates the performance of the EnKF-SQ. Accordingly, in what follows we set $\alpha = 1$. 

\begin{figure}[h]
\centering
\subfloat[\label{fig:initial_mean_thickness}]{\includegraphics[scale=0.2]{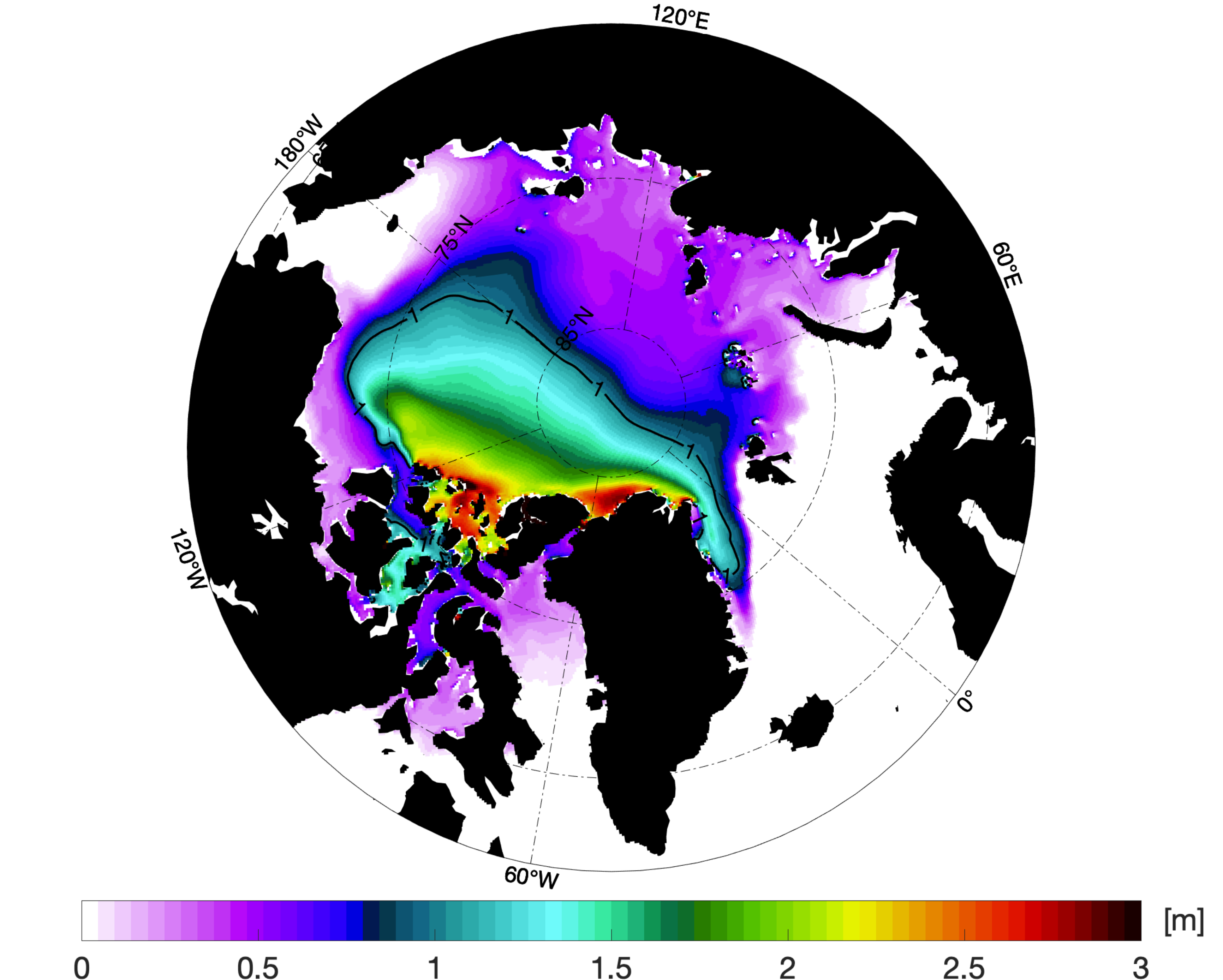}}
\subfloat[\label{fig:first_inc_enkfsq}]{\includegraphics[scale=0.2]{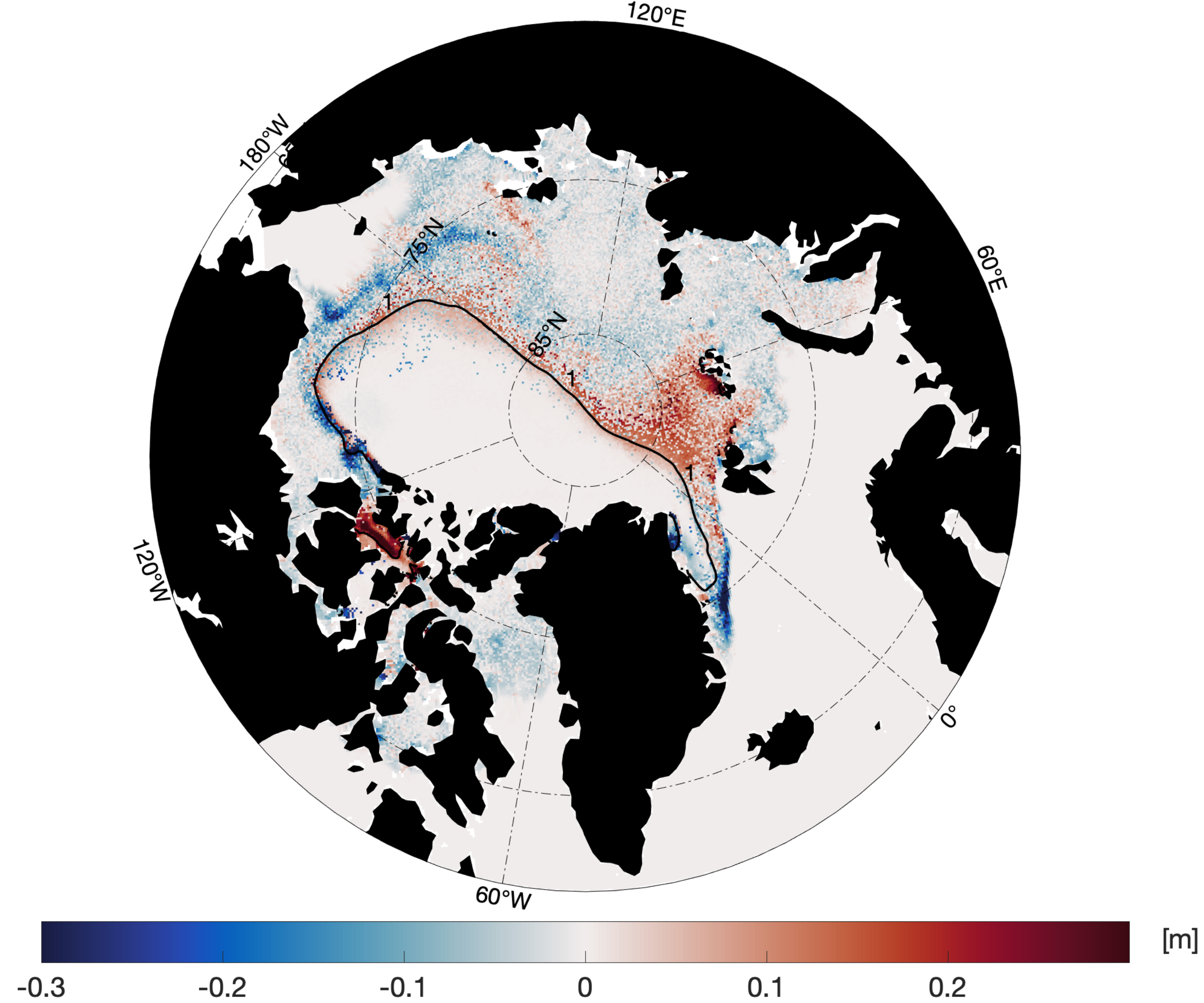}}
\caption{(a) Prior ensemble mean of the ice thickness on $11$ November $2014$. The solid black line is  the $1$ m SIT isoline. (b) The increment (analysis-forecast) for SIT after incorporating the observations.}\label{fig:Forecast_nd_update}
\end{figure}

To illustrate how the EnKF-SQ updates the SIT by assimilating range-limited SIT observations, we plot the prior mean (Figure \ref{fig:initial_mean_thickness}) and analysis increment (Figure \ref{fig:first_inc_enkfsq}) on $11$ November $2014$. The solid black line on both maps is the isoline for $1$ m of SIT. The forecast places the thick ice (up to $3$ m) north of Greenland and north-eastern part of Canada. The increments are not only visible outside of the $1$ m isoline but also inside the central Arctic region where only soft data are assimilated. It is important to notice that there is nearly zero increment in the central Arctic region and the Beaufort sea where the sea ice is thicker than $1.5$ m. This is because the EnKF-SQ analysis do not impose strong updates on the prior if it is above the detection limit and observations are out-of-range.

\subsection{Performance Assessment} \label{subsec:performance_assess}
\begin{figure}[h]
\includegraphics[width=\textwidth,height=9cm]{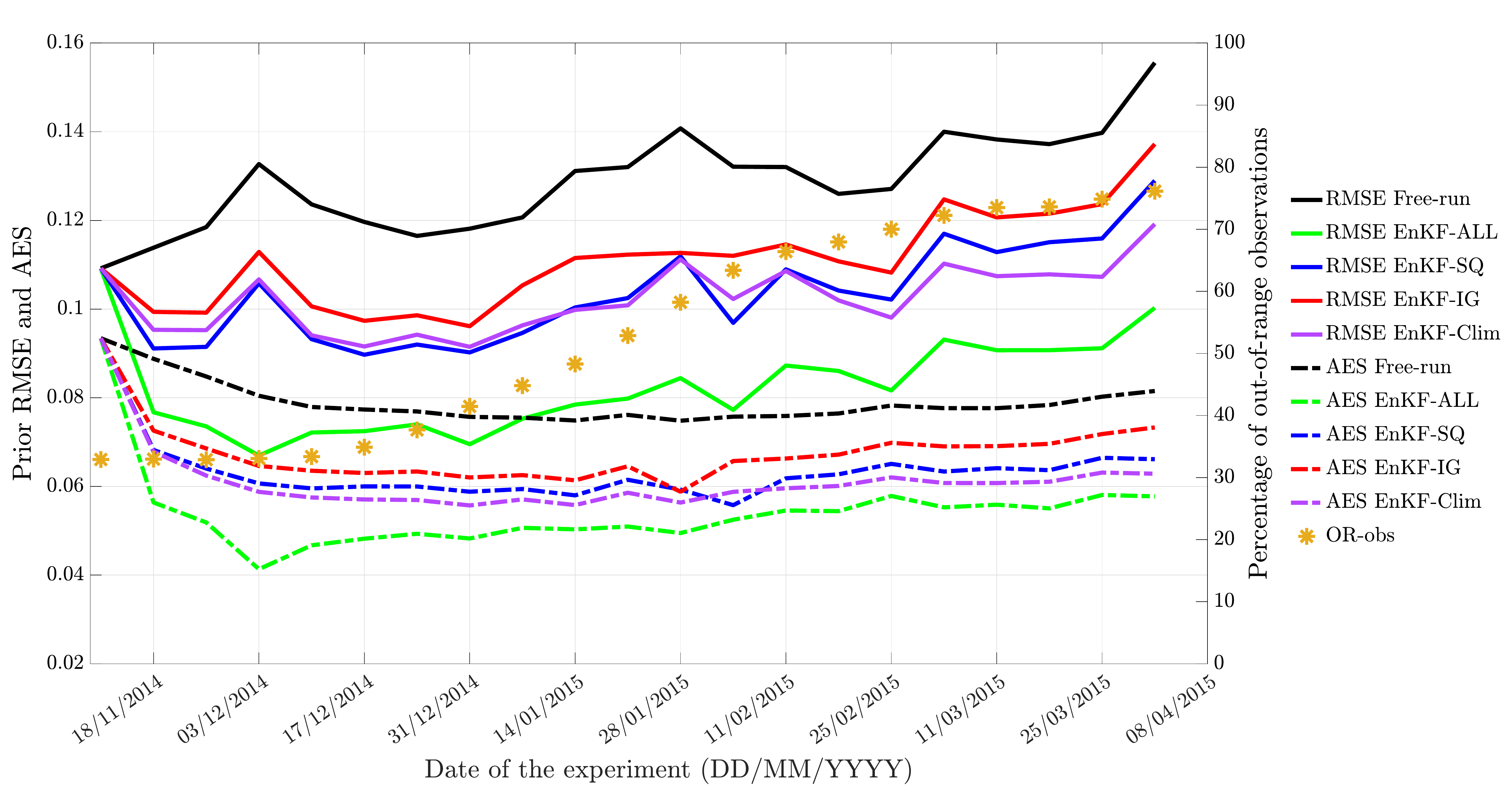}
\begin{centering}
\caption{Left y-axis: Time evolution of the prior RMSE (solid lines) and AES (dashed lines) for SIT estimates. Right y-axis: The orange asterisks represent the percentage of the out-of-range observations during assimilation resulting from the EnKF-SQ, EnKF-CLIM and EnKF-IG. \label{fig:All_schemes_RMSE_Spread}}
\end{centering}
\end{figure}

Figure \ref{fig:All_schemes_RMSE_Spread} shows the time evolution of the RMSE and AES of the prior SIT estimates obtained using the EnKF-ALL, EnKF-SQ, EnKF-CLIM, EnKF-IG and the Free-run. The percentage of OR observations (to the total number of observations) available at every cycle is added to the plot. As expected, EnKF-ALL outperforms all other schemes while EnKF-IG is the least accurate. It should be noted that there is an increasing trend in the RMSE, which is seasonally driven; a similar behavior reported in \cite{Xie2016}. Assimilating soft data with the EnKF-SQ clearly improves the prior RMSE compared to the EnKF-IG. This is consistent over the entire assimilation period. The number of OR observations gradually increases as the cold season intensifies leaving only a few hard data during the months of February and March 2015. Even with a very limited number of hard data, the EnKF-SQ outperforms EnKF-IG. The RMSE resulting from the EnKF-CLIM is marginally higher than that of the EnKF-SQ, except during the last three months of the assimilation experiment. The reason for this could be twofold: (i) In the early stages of the experiment, the climatology tends to overestimate SIT due to the large seasonal cycle compared to later months. This causes the climatology to pull the update towards large values and hence degrades the performance of the EnKF-CLIM. (ii) Fewer hard data leads to larger RMSE values in the EnKF-SQ as can be seen towards the end of winter and start of the spring. Overall, the RMSE and AES show consistent ensemble statistics such that sufficient variability is preserved in the system after cycling over time. 

\begin{figure}[h]
\includegraphics[scale=0.3]{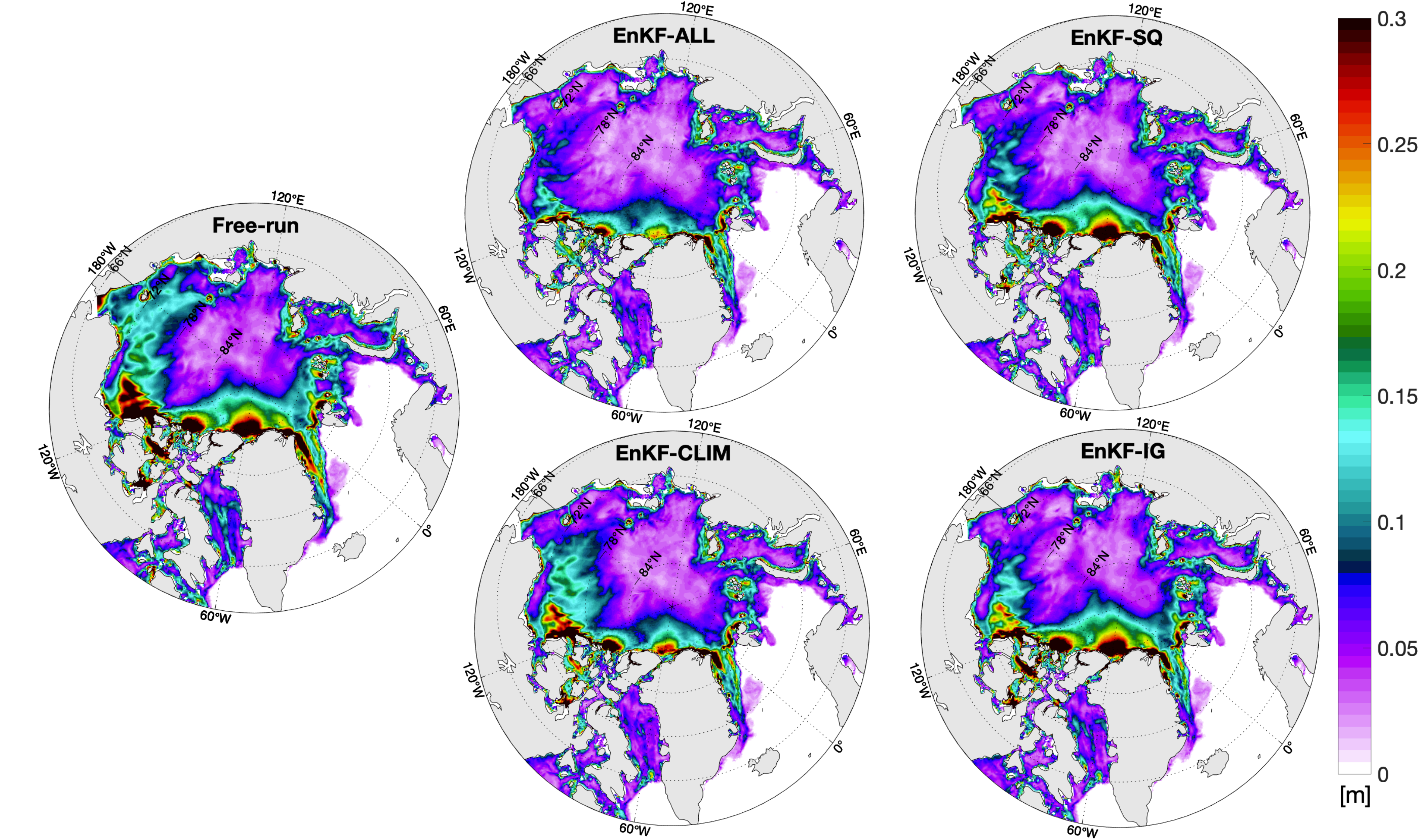}
\begin{centering}
\caption{Maps of time-averaged prior RMSE for SIT obtained using: EnKF-ALL (top left), EnKF-SQ (top right), Free-run (center left), EnKF-CLIM (bottom left) and EnKF-IG (bottom right). Averaging is done over the period of experiment, i.e., from November $2014$ to March $2015$. \label{fig:RMSE_map}}
\end{centering}
\end{figure}

In order to visualize area-wise improvements, we plot the map of time-averaged RMSE of the SIT prior estimates in Figure \ref{fig:RMSE_map}. The EnKF-ALL yields the best RMSE throughout the entire region. Compared to the EnKF-IG, the EnKF-SQ performs better in the central Arctic region, Greenland's north-eastern shelf, the Canadian Arctic Archipelago and in the Beaufort Sea. On average, the EnKF-SQ and EnKF-CLIM estimates are approximately $8$\% more accurate than those of the EnKF-IG.

The EnKF-CLIM, seems to produce larger improvements than the EnKF-SQ specifically along the Ellesmere island. However, it also increases the prediction error in the Beaufort sea more than that of the EnKF-IG. A number of reasons may explain this behavior. The climatology being too high compared to the seasonal mean yields an artificial increase of the model thickness, which happens to agree with the truth along the Ellesmere island. The recurrent update due to the assimilation of climatology is propagated dynamically by the Beaufort gyre into the Beaufort sea creating an anomaly compared to the truth, which is not thicker. 

\begin{figure}[h]
\begin{center}
\includegraphics[scale=0.275]{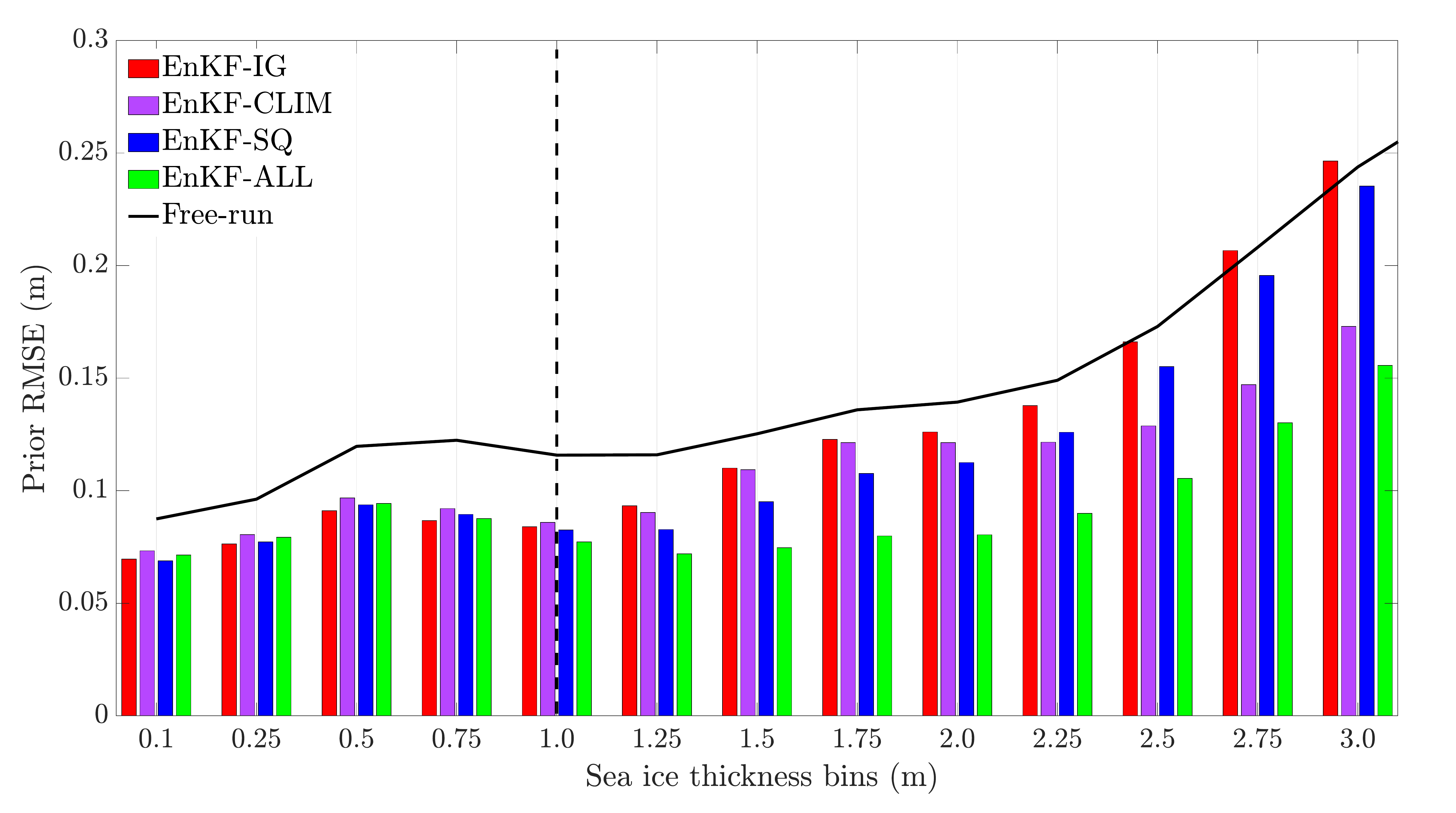}
\caption{Bar chart of time-averaged conditional prior RMSEs for SIT obtained using all tested DA schemes. Solid black line represents time averaged Free-run RMSE. Black dashed line depicts the $1$ m detection limit. The x-axis denotes the SIT bins with bin size of $25$ cm. The values on x-axis are the upper bounds of the SIT for that particular bin. }\label{fig:Bar_chart_RMSE}
\end{center}
\end{figure}

The analysis algorithm of the EnKF-SQ is designed such that improvements are expected mostly where SIT is close to the threshold. As a way to examine this, we computed the time-averaged RMSE of the prior SIT estimates for different ice thickness intervals of $25$ cm using all DA schemes (Figure \ref{fig:Bar_chart_RMSE}). The values on the x-axis of Figure \ref{fig:Bar_chart_RMSE} represent the upper bounds of each $25$ cm SIT bin interval except for the first bin of size $10$ cm because of the model $10$ cm minimum thickness. The RMSE for all DA schemes within each SIT bin is computed by finding the location of grid cells for which the observations fall within the bin interval. 

Figure \ref{fig:Bar_chart_RMSE} suggests that RMSE values for all schemes below $1$ m of SIT are approximately the same, as they all assimilate hard data. Once SIT increases beyond the detection limit, EnKF-ALL becomes the most accurate followed by the EnKF-SQ up to SIT of $2$ m. The EnKF-SQ performs as expected for observation values in the vicinity of the detection limit where the assimilation of soft data is clearly enhancing the accuracy compared to the EnKF-IG and EnKF-CLIM. The performance of the EnKF-SQ is not as good as the EnKF-CLIM for thicker ice, which can also be seen in Figure \ref{fig:RMSE_map} around the northern coast of Greenland. It is worth noticing that even though there is no data to assimilate for SIT $> 1$ m in the EnKF-IG scheme, it is performing better than the Free-run up to $2.25$ m of SIT. This advantage has been previously reported by \citet[see Figure 8]{Xie2016} and can be either due to the reduction of the positive bias in the free run (shown in Figure \ref{fig:Bar_chart_BIAS}) by assimilating thin ice only or due to dynamical model adjustments after assimilation. In other words, improvements to thin ice are propagated in time to the period where ice gets thicker. 

\subsection{Bias and Skewness Analysis} \label{subsec:bias_skewness}
\begin{figure}[h]
\begin{center}
\includegraphics[scale=0.275]{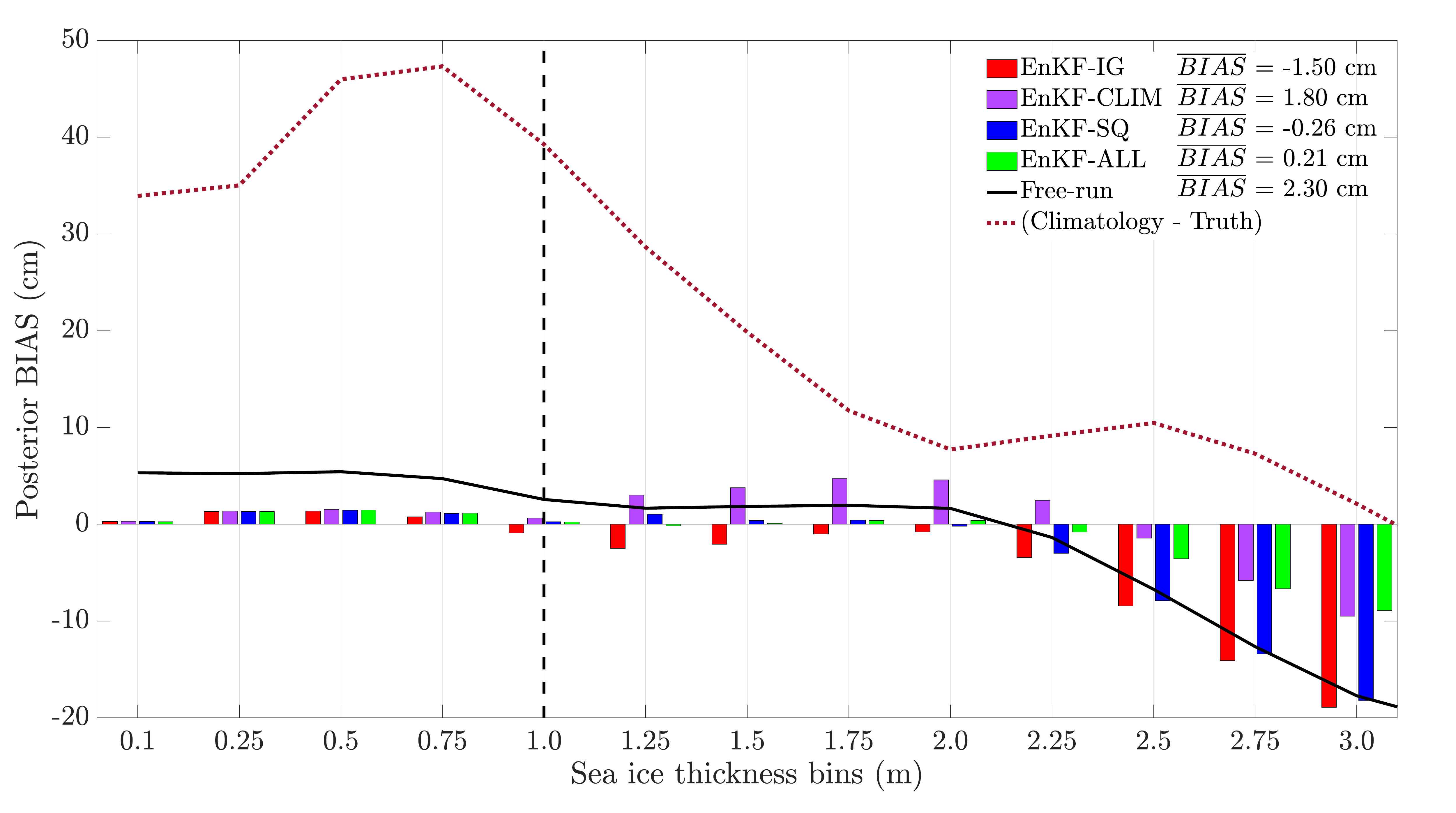}
\caption{Bar chart of time-averaged posterior bias for SIT obtained from all tested DA schemes. Solid black line represents the time-averaged bias for SIT obtained using the Free-run. Red dotted line represents the time-averaged difference of climatology and truth. SIT bins are displayed on the x-axis with a bin size of $25$ cm. Reported in the legend are the time-averaged-weighted total mean bias including the bins for ice thicker than $3$ m, which is not shown here. The weights are computed as a fraction of the number of grid cells falling in specific bin interval over total number of grid cells.} \label{fig:Bar_chart_BIAS}
\end{center}
\end{figure}

The EnKF-IG updates the prior members by only assimilating observations of  thin ice with a maximum thickness of $1$ m. This causes the algorithm to introduce negative conditional bias for thick ice (knowing that the observation is thin ice, the assimilation reduces the ice thickness more that it can thicken it). Similarly, the EnKF-SQ update may introduce a bias towards the detection limit due to assimilation of soft data and the EnKF-CLIM towards the climatology. To investigate these likely biases in different DA schemes, we present a bar chart of time-averaged conditional bias for the posterior estimates of SIT in Figure \ref{fig:Bar_chart_BIAS}. The conditional bias is calculated by finding the location of the grid cells for which the observations fall within the SIT bin interval. The positive values represent an overestimation of SIT after the assimilation and vice versa. 

The four DA runs exhibit a small negligible positive bias of approximately $0.5$ to $1$ cm for thin ice. The Free-run bias, on the other hand, is larger than $\sim6$ cm. Above the threshold limit, there is a clear positive bias of $5$ to $7$ cm in the EnKF-CLIM posterior estimates, up until $2$ m. As seen earlier, the climatology tends to overestimate the truth during the first few months of the experiment when the ice is thin (red dotted line in the Figure \ref{fig:Bar_chart_BIAS}). EnKF-IG estimates, over the same interval, exhibit a small negative bias, possibly left over from the conditional assimilation of thin ice. It is important to note that there is almost zero bias in the EnKF-SQ estimates, matching that of the EnKF-ALL for $1 \leq \text{SIT} \leq 2$ m.

There is a systematic increasing negative bias for SIT $> 2$ m, which reaches almost $20$ cm for SIT $= 3$ m in the Free-run, EnKF-IG and EnKF-SQ. A similar trend of negative bias is also observed in the EnKF-ALL and EnKF-CLIM runs but to a slightly lesser extent. The negative bias in the Free-Run is likely due to the perturbation of the forcing fields, specifically the wind perturbations, which can cause erratic movements of ice that export thicker sea ice into areas of thinner ice. Since all assimilation runs use perturbed winds, this effect is likely to impact the EnKF-IG and EnKF-SQ more than the EnKF-ALL and EnKF-CLIM. In addition, it is important to mention that there are fewer grid points (not shown here) in the bins for thicker ice compared to thin ice, which may also affect the estimation of the bias for these bins, making them statistically less significant.

\begin{figure}[h]
\begin{center}
\includegraphics[scale=0.275]{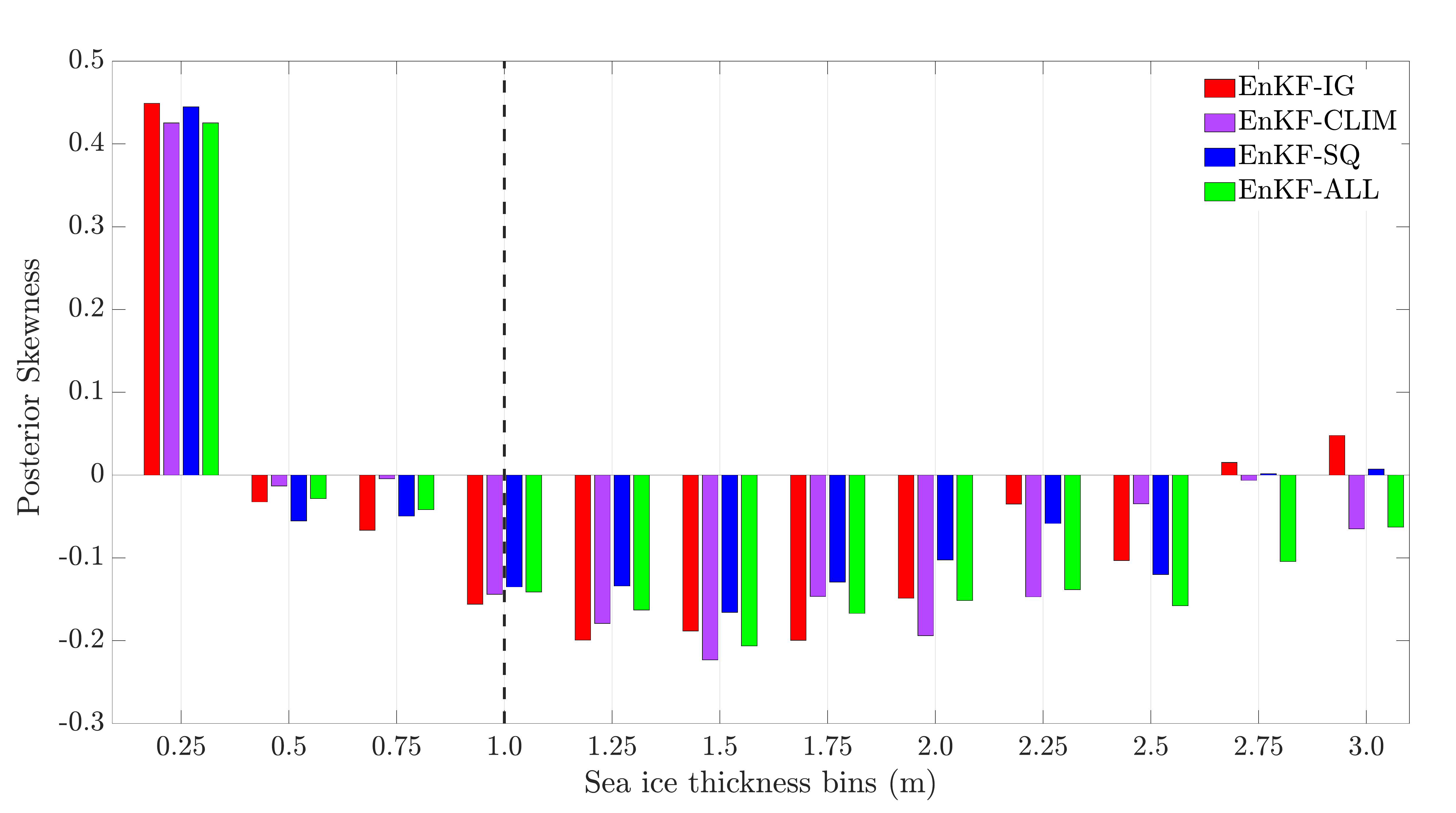}
\caption{Bar chart of the conditional posterior skewness for SIT estimates obtained using all tested DA schemes and computed at the final assimilation time. Dashed black line represents the detection limit of $1$ m SIT.} \label{fig:skewness}
\end{center}
\end{figure}

As discussed in \cite{Shah2018}, the two-piece Gaussian observation likelihood may influence the shape of the posterior distribution, making it skewed and thus less Gaussian. In order to examine this, we evaluate and plot the conditional skewness of the posterior estimates of SIT only at the last assimilation step in Figure \ref{fig:skewness}. The conditional skewness of the posterior is calculated as the average value of the skewness for all grid cells where the truth falls within the interval of a bin in consideration. Note that contrary to the computation of the conditional BIAS at the location of the observations, the conditional skewness is computed at the truth locations. 

As shown in the figure, thin ice (SIT $\leq25$ cm) yields noticeable skewness in the posterior estimates for all schemes. In the first bin, the truth is close to zero meters (open water) and hence all instances where thin ice has melted in the assimilation run count as zero value. On the other hand, freezing instances lead to various thickness values above $25$ cm. Both effect together can make the distribution skewed. The bin between zero and $10$ cm shows even larger skewness and has been removed for a better visual presentation. Other than the first bin, a small negative skewness is observed for all the schemes. One possible explanation is the fast melting of ice, drifting over warm waters; a situation enhanced by the lack of coupling with the ocean in the assimilation. This result confirms that the EnKF-SQ, although it uses a skew 2-piece Gaussian likelihood, does not introduce any noticeable positive skewness in its posterior.

\subsection{Physical Consistency}
Ice-ocean models are essential tools for computing integrated quantities that are often difficult to estimate from observations only. Sea ice volume and water transport between ocean basins are such high interest quantities for climate studies. Therefore, it is important to evaluate these quantities to verify that the use of data assimilation does not cause physical inconsistencies.

\begin{figure}[h]
\centering
\subfloat[\label{fig:daily_sea_ice_volume}]{\includegraphics[width=0.5\textwidth, height=6.3cm]{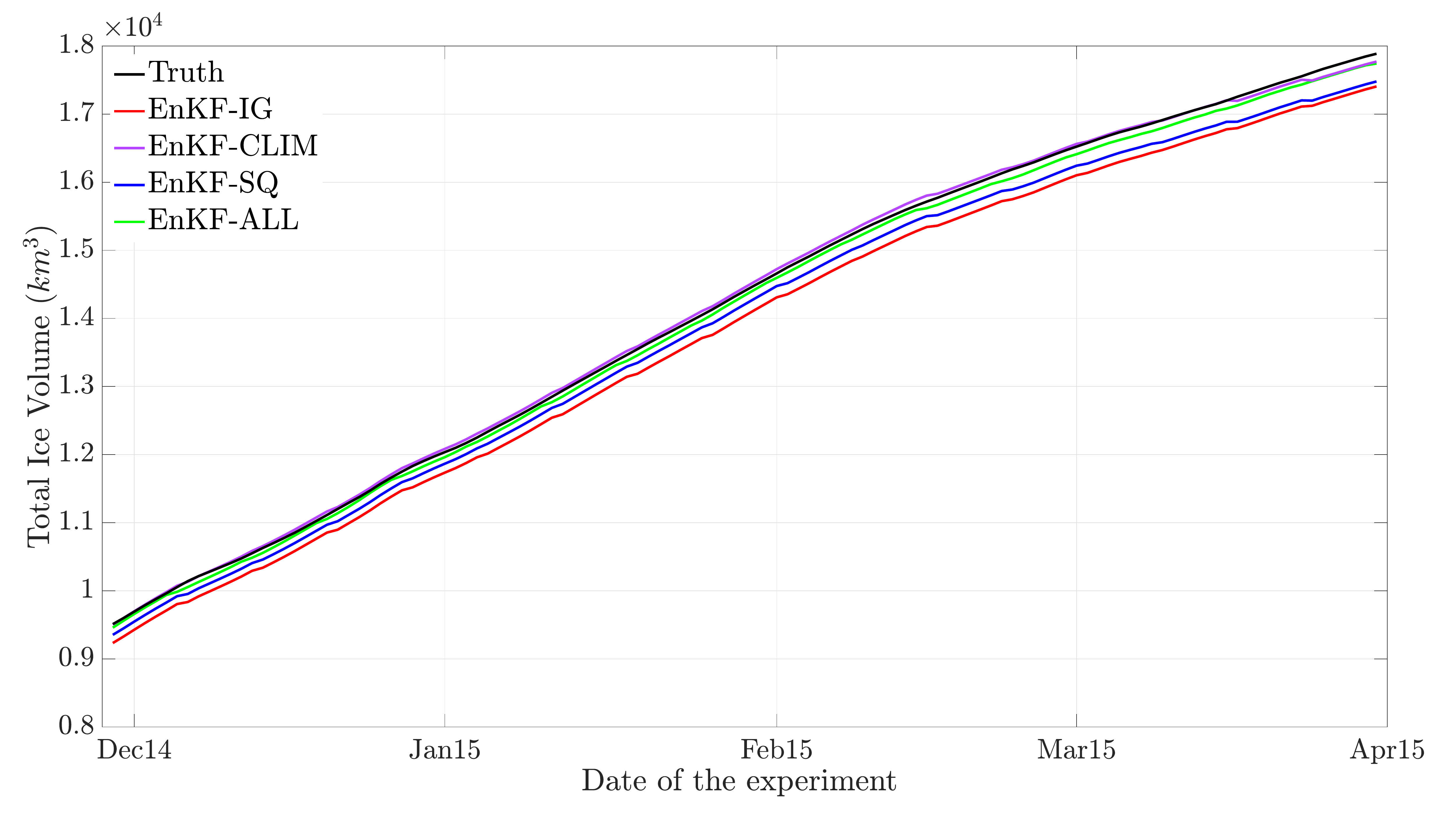}}
\subfloat[\label{fig:diff_ice_volume}]{\includegraphics[width=0.5\textwidth, height=6.3cm]{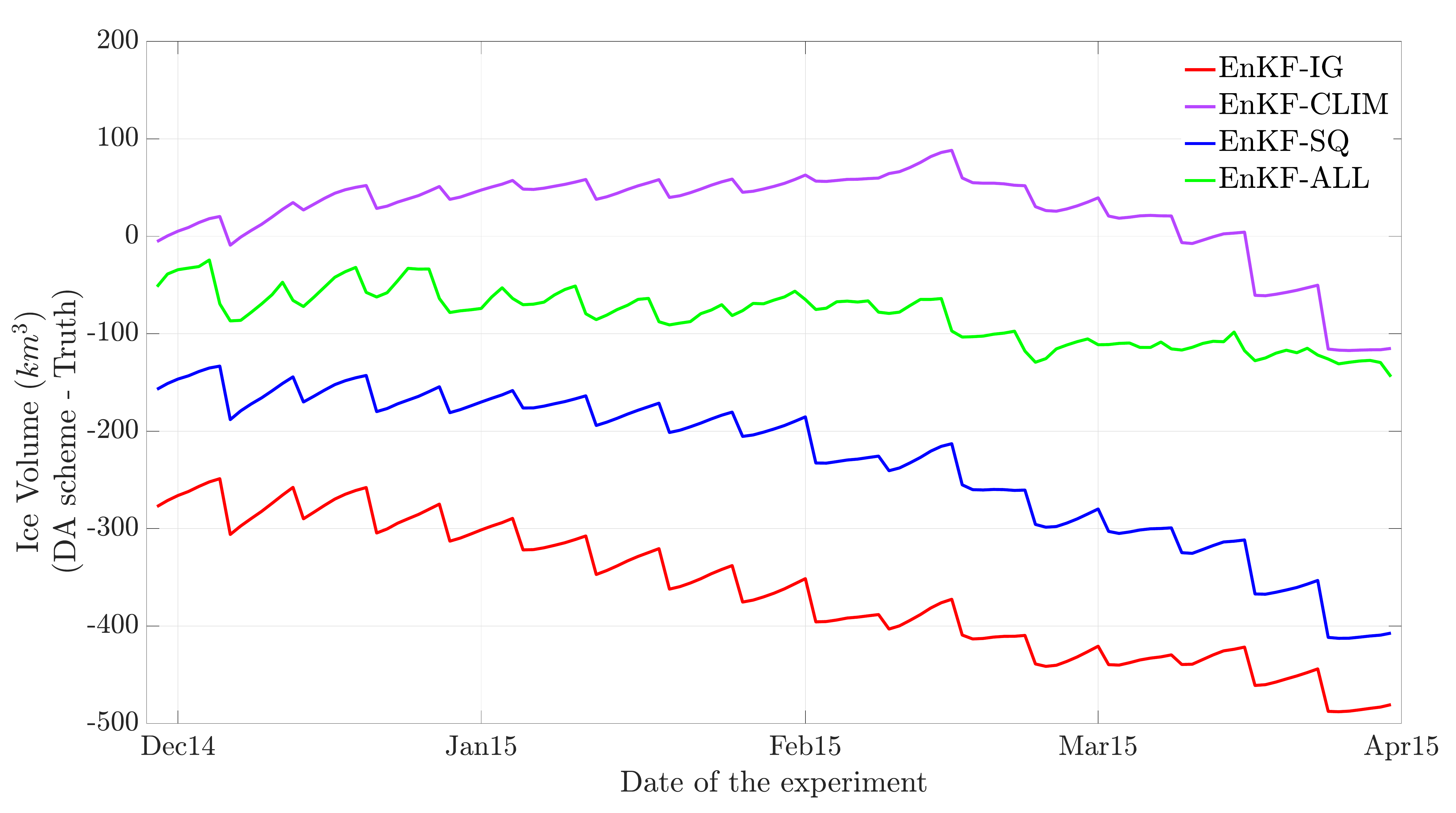}}
\caption{(a) Daily ensemble average of sea ice volume over the TOPAZ4 model area for the entire experiment time. (b) Difference of sea ice volume from the truth and all tested DA schemes. }\label{fig:ice_volume}
\end{figure}

The total sea ice volume is the integral of sea ice concentration times the sea ice thickness over the entire model area. Its evolution for the different assimilation runs is shown in Figure \ref{fig:daily_sea_ice_volume}. The difference between the assimilation runs compared to the true sea ice volume (Figure \ref{fig:diff_ice_volume}) is relatively small. This is because none of the DA schemes has extensively added or removed ice during the assimilation run. In Figure \ref{fig:diff_ice_volume} a classical seesaw Kalman update behavior is observed. The comparison also reveals that most methods tend to underestimate the ice volume except for EnKF-CLIM. 

As described earlier, the EnKF-IG has a negative SIT bias, which translates to a nominal loss of between $300$ km$^3$ to $500$ km$^3$ of sea ice volume from the beginning to the end of the winter (less than $3\%$ of the total simulated ice volume). Seesaw of the time series curves confirm that the EnKF-IG update does remove some ice, which grows back during the subsequent TOPAZ4 model run. The EnKF-SQ does only partially mitigate this loss by $100$ to $200$ km$^3$ of ice. Surprisingly, the EnKF-ALL is not bias-free either with a loss of up to $100$ km$^3$ of ice, which can be caused by various sub-optimal aspects of the data assimilation system, in particular the aforementioned effect of wind perturbations on the areas of thickest ice and the weakly coupled DA. These effects also contribute to the low bias in the other two methods. 

The EnKF-CLIM ice volume is closest to the truth run with a little overestimation in the beginning of the winter, then an underestimation in the spring. The construction of the climatological data can explain this trend: since SIT data above one meter only have been retained, the climatology overestimates the SIT in the beginning of the winter but then underestimates the SIT in the midst of the winter because it also accounts for summer SIT. A different construction of the SIT climatology data would have led to a different tendency in EnKF-CLIM. 

\section{Discussion and Conclusions} \label{sec:conclusion}

The purpose of this paper is to demonstrate the usefulness of assimilating range-limited observations with the new EnKF-SQ DA scheme under a realistic  experimental setup. Compared to the stochastic EnKF, the main algorithmic difference is the need to compute a different Kalman gain for each ensemble member, depending on the location of the member to the threshold when the observation is out-of-range. This does not make the EnKF-SQ less efficient, but rather prevents the algorithm from being included as a simple extension of existing EnKF codes: it cannot be expressed with an ensemble transform matrix.  

The assimilation of synthetic sea ice thickness data with a upper detection limit of $1$ m in a coupled ice-ocean model of TOPAZ4 is demonstrated using the EnKF-SQ and shown to have a useful impact on SIT estimates. The results obtained with SMOS-like observations can be generalized to CryoSat-2 like observations by reversing the upper limit into a lower limit. Thus, merging the two products may not be necessary because each satellite data can be assimilated in a separate EnKF-SQ step.

Different assimilation experiments are conducted to assess the performance of the EnKF-SQ against other EnKF configurations assimilating only thin ice; both thin and thick ice; and climatology during a winter period in the Arctic. The study shows that assimilating soft data improves the forecast accuracy compared to ignoring them by approximately $8$\%, particularly where sea ice approaches the detection limit. Such a difference can be important in the performance of an operational system.

The performance exhibited by assimilating a reasonably accurate climatology was similar to the EnKF-SQ. Also, our choice of climatology being annual rather than seasonal may explain some of the flaws in the EnKF-CLIM. Nonetheless, the context of twin experiments is very favorable to EnKF-CLIM because the climatological truth is perfectly known; a case which is not true in realistic situations. For instance, in summer there are very few ice thickness measurements and thus it is difficult to construct a meaningful climatology. To this end, it is essential to investigate and compare the performance of the EnKF-SQ and EnKF-CLIM in a context of a biased model twin experiment and with a range of toy models (from linear to non linear regimes).

Assessing the bias of the analysis showed that there is no introduction of any significant bias by the EnKF-SQ, other than the negative bias for thicker ice which is observed in all tested DA schemes. Likewise, the posterior distributions resulting from the application of the EnKF-SQ did not consist of any noticeable higher order moments that could result in undesirable non-Gaussian features because of the two-piece Gaussian likelihood. This is most likely the case for all realistic applications where one would expect relatively small assimilation updates coming regularly in time. Furthermore, the choice of out-of-range (OR) observation error variance was not found to be very critical. A wide range of values for this parameter were tested and lead to acceptable performance of the EnKF-SQ. Ways of estimating $\sigma_{or}^2$ adaptively in space and time is currently being investigated and will be reported in a follow-up study. Concerning the physical constraints of the model, the EnKF-SQ estimates were found to be physically consistent and comparable to other tested assimilation schemes.

The EnKF-SQ therefore makes a viable data assimilation strategy for range-limited observations in high-dimensional nonlinear systems. Future research will focus on assimilating real data, in which the EnKF-SQ is confronted with large observation biases unlike the presented twin experiments setup.

\section*{Acknowledgements}
The research is funded by the Nordic Center of Excellence EmblA (Ensemble-based data assimilation for environmental monitoring and prediction) under NordForsk contract number 56801. The Copernicus Marine Services and the Nansen Scientific Society have also contributed to the funding. Norwegian grants of the computer time (nn2993k) and data storage space (ns2993k) have also been used for the simulations.

\bibliographystyle{wileyqj}

\end{document}